# Are we living in a quantum world?
# Bohr and quantum fundamentalism[*]

Henrik Zinkernagel
Department of Philosophy I
University of Granada, Spain.
zink@ugr.es

**Abstract**
The spectacular successes of quantum physics have made it a commonplace to assert that we live in a quantum world. This idea seems to imply a kind of "quantum fundamentalism" according to which everything in the universe (if not the universe as a whole) is fundamentally of a quantum nature and ultimately describable in quantum-mechanical terms. Bohr's conception of quantum mechanics has traditionally been seen as opposed to such a view, not least because of his insistence on the necessity of the concepts of classical physics in the account of quantum phenomena. Recently, however, a consensus seems to be emerging among careful commentators on Bohr to the effect that he, after all, did subscribe to some version of quantum fundamentalism. Against this consensus, and by re-examining the historical record, I will defend a variant of the traditional reading of Bohr in which (1) the answer to what an object is (quantum or classical) depends on the experimental context; and (2) in principle, any physical system can be treated quantum-mechanically but not all systems can be treated that way at the same time.

## 1. Introduction

Niels Bohr's 1913 atomic model was a revolutionary step in a development which culminated in the mid-1920s with the theory of quantum mechanics. Ever since then, physicists and philosophers have been discussing what the theory means for our understanding of the world. As is well known, Bohr contributed substantially to this debate – not least through his famous discussions with Einstein. And yet, even among careful Bohr scholars, disagreements remain about what Bohr's view really was. This paper is part of a series that aim to clarify and defend Bohr's interpretation of quantum mechanics.[1]

Quantum mechanics is an immensely successful theory. Not only have all its predictions been experimentally confirmed to an unprecedented level of accuracy, allowing for a detailed understanding of the atomic and subatomic aspects of matter; the theory also lies at the heart of many of the technological advances shaping modern society – not least the transistor and therefore all of the electronic equipment that surrounds us.

Against this background, it is hardly surprising that many are attracted to the idea that quantum mechanics, or some of its successors like quantum field theory or string theory, is a universal and fundamental theory of the world. Thus, what could, somewhat provocatively, be called 'quantum fundamentalism' is today a widespread view among both physicists and philosophers. The view can more accurately be characterized as follows:

> Quantum Fundamentalism
> Everything in the universe (if not the universe as a whole) is fundamentally of a quantum nature and ultimately describable in quantum-mechanical terms.

---

[*] Published in F. Aaserud and H. Kragh (eds.), *One hundred years of the Bohr atom: Proceedings from a conference*. Scientia Danica. Series M: Mathematica et physica, vol. 1. Copenhagen: Royal Danish Academy of Sciences and Letters, 2015, pp. 419-434.
[1] Several points (especially in sections 3 and 5) of the present paper, which is subject to length constraints of the conference proceedings, are treated more extensively in Zinkernagel (2016).

In this formulation, quantum fundamentalism contains both an ontological and an epistemological thesis: that everything *is* of a quantum nature is an ontological claim, whereas the idea that everything can (at least in principle) be *described* in quantum terms is epistemological. The ontological component of quantum fundamentalism can also be expressed as the idea that we live in a quantum world.

At least on the face of it, Niels Bohr's conception of quantum mechanics stands in contrast to quantum fundamentalism. Thus, Bohr insisted on the necessity of the concepts of classical physics in the description of quantum phenomena, e.g.,[2]

> [I]t is decisive to recognize that, *however far the phenomena transcend the scope of classical physical explanation, the account of all evidence must be expressed in classical terms.* (Emphasis in original)

In spite of such statements, a recent consensus seems to be emerging among careful commentators on Bohr to the effect that he, after all, did subscribe to some version of quantum fundamentalism. Against this, I argue below that Bohr is best understood as *not* being a quantum fundamentalist – neither in the ontological nor in the epistemological sense. But first I will briefly outline a key problem for quantum fundamentalism which will be important for the discussion to follow.

## 2. The quantum measurement problem

For all its impressive predictive and practical successes, quantum mechanics cannot account for the most basic fact of experiments – namely that they have definite outcomes. This is known in the literature as the quantum measurement problem. Stated briefly, the problem is that *if* everything, including measurement apparatuses, is quantum (and correctly described by Schrödinger's equation), then there ought to be superpositions in the measurement outcomes. For our purposes, superpositions may be taken to roughly correspond to objects being at different places at the same time. Such superpositions are essential at the microscopic level, e.g., in the explanation of the interference pattern seen in the famous double slit experiment. But if everything is quantum then superpositions at the microscopic level become "amplified" to the macro level, and there should thus, for instance, be apparatus pointers in various positions at the same time. Clearly this contradicts ordinary experience.

The measurement problem has been an important motivation for a number of different interpretations of quantum mechanics, although none of these are widely believed to be free of problems.[3] Thus, responses to the problem have been sought by invoking assumptions – such as observer-induced collapse, many worlds, hidden variables, or modified dynamics – which clearly or arguably go beyond the quantum formalism itself to somehow 'explain away' why no quantum strangeness is seen in the measurement results.

---
[2] Bohr (1949), p. 39.
[3] For a simplified description of the measurement problem, a brief characterization of the main proposals for its solution, and its relation to what is known as decoherence, see Zinkernagel (2011), section 2.1.

As noted above, Bohr's view of quantum mechanics has traditionally been seen as opposed to quantum fundamentalism, not least because he insisted on a classical (not quantum) description of the measurement apparatus. This amounts to dissolving the measurement problem from the outset, since the apparatus and its pointer should not be thought of as being in quantum superpositions. Thus, the apparatus and its pointer are always in definite positions. However, Bohr's idea seems to be problematic as it suggests that there is a separate classical realm of reality. And it is not at all clear how one could divide up the world in a classical and a quantum part. This is a main reason why recent studies have questioned whether the traditional view of Bohr is correct. Conversely, the question is whether Bohr can reasonably be associated with quantum fundamentalism. This is the question to be addressed in the following two sections.

## 3. Bohr as epistemological and ontological quantum fundamentalist?

The literature has witnessed many interpretations of Niels Bohr's thoughts. Often, Bohr has been associated with specific philosophical schools such as positivism, Kantianism, or pragmatism. But at least since Don Howard's interesting and influential analysis of Bohr's understanding of classical descriptions from 1994,[4] there has been a tendency among commentators to try to understand Bohr more independently of traditional philosophical positions.

In what he calls a reconstruction of Bohr, Howard argues that a classical description is best understood as a description in which one denies the entanglement between, e.g., a measurement apparatus and an electron. But entanglement, which means that two interacting systems cannot be regarded as two separate systems with mutually independent properties, is a quantum mechanical fact. In order to account for the observed definite properties of the apparatus, one should therefore proceed *as if* the apparatus and the electron were not entangled. The technical implementation of Howard's idea consists in replacing entangled quantum states with so-called mixed quantum states appropriate to the given experimental context.[5] Howard's mixed state is *separable*, which formally means that it can be written as a product of the wave functions of the electron and the measuring apparatus. The importance of this lies in the fact that such a separable mixed state is compatible with both the electron and the apparatus' pointer being in a well-defined (and not super-) position. In this way, for Howard's Bohr, classical descriptions are not descriptions by classical physics, but rather a (context-dependent) denial of quantum entanglement between apparatus and object. But since all systems are still described by quantum states, Howard claims that:[6]

> The distinction between classical and quantum modes of description … is implicit in quantum mechanics itself, and thus not a mark of some fundamental ontological or epistemological distinction.

---

[4] Howard (1994); see also Howard (2004).
[5] Howard (1994), p. 203 ff.
[6] Howard (1994), p. 204.

However, as far as I can see, there are major problems associated with Howard's reconstruction. Both because the reconstruction seems inconsistent with several central points in Bohr's texts, and because it ultimately does not solve the measurement problem. One immediate problem is that even a separable (non-entangled) quantum state is still a quantum state. More specifically, this means that even if the entangled state is replaced by a product state (a product of the wave functions of the object and the apparatus respectively), the apparatus is still ascribed a quantum mechanical wave function. But the attribution of a wave function to the apparatus seems incompatible with Bohr's idea,[7] that a wave function can only be attributed to a (quantum) system in a context where another system (typically the measuring device) is omitted from the quantum description, and thus *not* represented by means of a wave function.[8]

In addition, a quantum (wave function) description of the measuring apparatus implies that although it, or its pointer, could in a given context be attributed a well-defined position, it could not simultaneously (due to Heisenberg's uncertainty relations) be attributed a well-defined momentum. This is inconsistent with Bohr's insistence on several occasions that the apparatus must be described "…in *purely* classical terms, excluding in principle any regard to the quantum of action" [my emphasis].[9]

A further problem with Howard's reconstruction is that the measurement problem remains unresolved. For, as also Schlosshauer and Camilleri have pointed out, Howard does not explain *why* one is allowed to replace the original entangled state with an appropriate mixed and separable state.[10] Thus one cannot simply conclude that the measuring instrument (or its pointer) is in a particular state and in a particular position. Against this, Howard might argue that Bohr would have to assume the replacement of the entangled state with an appropriate mixed state at the descriptive level. For without the associated separability between the measuring apparatus and the object, it would not be possible from the formalism to explain why measurements have results at all. But, as we shall see below, separability at the descriptive (epistemological) level is not enough, as the measurement problem is ultimately a consequence of the ontological component of quantum fundamentalism.

## 4. Bohr as ontological (but not epistemological) quantum fundamentalist?

Klaas Landsman is another careful Bohr interpreter who sees Bohr as a quantum fundamentalist, albeit only in its ontological variant. In his view, which is similar to the view expressed by Schlosshauer and Camilleri,[11] Bohr's classical description of, e.g., a measurement apparatus simply means that it should be treated *as if* it were a classical physical system. According to Landsman, this implies that all measurable quantities of the apparatus can be regarded as having definite values after the measurement.[12] Landsman summarizes his position thus:[13]

---

[7] See, e.g., Bohr (1958), p. 5.
[8] However, I think Howard's reading of Bohr is viable and illuminating if wave functions are ascribed only to *parts* of the apparatus (as opposed to the apparatus as a *whole*), see Zinkernagel (2016).
[9] Bohr (1958), p. 4.
[10] Schlosshauer and Camilleri (2008), p. 23.
[11] Schlosshauer and Camilleri (2008).
[12] See Landsman (2007), p. 438. Landsman's formal ideas regarding how to understand Bohr are developed in Landsman (2006).
[13] Landsman (2007), p. 437.

> Bohr … believed in the fundamental and universal nature of quantum mechanics, and saw the classical description of the apparatus as a *purely epistemological move without any counterpart in ontology*, expressing the fact that a given *quantum* system is *being used* as a measuring device. [Emphasis in original]

In my view, Landsman is right with regard to what it means for Bohr to describe something as classical. However, I think Landsman's idea of Bohr as an ontological (but not epistemological) quantum fundamentalist faces several problems. For a start, it is not obvious that Bohr would have agreed with the distinction between the epistemological and ontological level regarding quantum fundamentalism. Take for example this oft-quoted remark by Aage Petersen:[14]

> When asked whether the algorithm of quantum mechanics could be considered as somehow mirroring an underlying quantum world, Bohr would answer, "There is no quantum world. There is only an abstract quantum physical description. It is wrong to think that the task of physics is to find out how nature *is*. Physics concerns what we can *say* about nature. [Emphasis in original]

As pointed out by Favrholdt and Hebor,[15] this is *not* a rejection of the existence of objects, such as electrons or atoms, which are typically described quantum mechanically. Rather Bohr's point is to reject the correspondence view of scientific theories; that is, the idea that the structures and objects treated in quantum theory directly reflect or correspond to how the world actually is. Such a refusal is a natural consequence of the *complementarity* view that Bohr introduced in 1927.

Complementarity means that the attribution of certain properties to quantum objects can take place only in experimental contexts that are mutually incompatible. Thus, for example, an experiment that can determine the position of an electron cannot be used to determine its velocity (or momentum). Complementary properties, such as position and momentum, are both necessary for a full understanding of the object but, as manifested in Heisenberg's uncertainty relations, the object cannot possibly be attributed precise values of both properties at the same time. Moreover, Bohr often pointed out that the quantum mechanical formalism is symbolic in that it cannot be given a direct pictorial or visualizable interpretation (which the correspondence view of theories suggests). To justify this, Bohr noted among other things that the quantum wave function in general "lives" in a mathematical space with more than the usual three dimensions.

In the quote by Petersen, Bohr emphasizes what one can *say* about nature, and rejects the question of how nature *is* in itself. This is consistent with the fact that Bohr almost always stresses epistemological rather than ontological aspects of quantum mechanics. But even if one for a moment ignores this, it is far from clear what, from Bohr's viewpoint, it would mean to assert that an object in itself *is* a quantum system. An obvious suggestion would be to think that the deep nature of objects is correctly represented by quantum wave functions. But Bohr could not have supported this idea. For in his view, the wave function is symbolic, and it can be attributed to objects *only* in experimental contexts which are not themselves described by such wave functions.

---

[14] Petersen (1963), p. 12.
[15] Favrholdt and Hebor (1999)

Another problem with the idea of Bohr as ontological quantum fundamentalist is that the measurement problem remains unresolved. For if the apparatus is essentially quantum, and if this means that its correct representation is ultimately through a wave function, then one cannot account for the fact that experiments have well-defined results. Indeed, it is precisely the unavoidable superpositions of wave functions that imply that measurements have no results. This consequence of quantum mechanics cannot be changed by considering the measurement apparatus as classical merely at the descriptive level.

## 5. Bohr's restricted quantum universalism

But how did Bohr then see the relationship between the classical and the quantum? To answer this, one must first ask two other questions: Why did Bohr insist on a classical description of the measurement apparatus? And what did he think of the measurement problem? In my view, one can find in Bohr's writings two often overlooked arguments for the necessity of classical descriptions which could be called, respectively, the *closedness* and *reference system* argument. The first of these also contains what I think must be Bohr's answer to the measurement problem.

Let me start with the measurement problem. In his published writings Bohr never discusses this problem, originally addressed by von Neumann in 1932,[16] in terms of superpositions of the states of the measurement apparatus. Nevertheless, Bohr was familiar with von Neumann's account, and I think it is most likely that Bohr was also aware that his own view constituted a solution, or rather dissolution, of the measurement problem. For example, Bohr says:[17]

> …every atomic phenomenon is closed in the sense that its observation is based on registrations obtained by means of suitable amplifications devices with irreversible functioning such as, for example, permanent marks on a photographic plate … [T]he quantum-mechanical formalism permits well-defined applications referring only to such closed phenomena.

Thus, the quantum mechanical formalism can only be applied to phenomena whose observation results in well-defined and irreversible outcomes (call this closedness). But the point of the measurement problem is precisely that a pure quantum mechanical treatment of both the apparatus and the studied object cannot give rise to such well-defined outcomes! So Bohr's point can be seen as a solution to (or dissolution of) the measurement problem: if the apparatus is classical, so that there is a well-defined pointer position at any time, then there is no problem with macroscopic superpositions at the end of measurements.

At first sight, this argument of closedness and the associated solution to the measurement problem seem to have an obvious difficulty. For the solution appears to require that the world is divided into an essentially classical and an essentially quantum mechanical part – roughly equivalent to the distinction between the macroscopic (e.g., a measuring apparatus) and the microscopic (e.g., an electron). As many authors have noted, such a division would be deeply problematic. Among other reasons, this is because it is then difficult to explain how macroscopic objects, such as measuring apparatuses or cats, could be composed of atoms which are indeed microscopic objects. And atoms seem to be paradigmatic examples of quantum objects, not least because quantum theory was developed precisely to explain atomic phenomena.

---

[16] von Neumann (1955), p. 417 (original in German 1932).
[17] Bohr (1954), p. 73.

In any case, Bohr did *not* think that there is such a fixed division of the world. Rather, Bohr emphasized that a macroscopic measuring apparatus, or at least parts of it, can also be treated quantum mechanically. However, it is only in special situations that a quantum treatment of a macroscopic object is relevant because quantum effects are in general insignificant for such objects. Indeed, Bohr often noted that quantum effects in a phenomenon are expressed in terms of the quantum of action. And such effects are generally too small to appear on a macroscopic or everyday scale, where typical actions are far greater than the quantum of action.

This idea suggests a sense in which Bohr might after all have agreed that objects are essentially quantum. For Bohr believed that the quantum of action (Planck's constant) is a basic empirical discovery that symbolizes an aspect of wholeness in atomic processes. Such processes cannot be separated from their interaction with the measuring device. Moreover, Bohr stresses that the classical description of objects is strictly speaking valid only as an approximation in the limit where one can ignore the quantum of action. This seems to imply that all objects are in some sense dependent on the quantum of action.

Does this then mean, as Landsman holds, that Bohr's requirement of a classical description of the measuring device is merely a pragmatic or epistemological arrangement, and that the device is really a quantum mechanical system? I think not. For the application of quantum mechanics to a system depends on disregarding – not just in practice but also in principle – the quantum of action for some other system. There are at least two reasons for this, and they have to do, respectively, with the already mentioned measurement problem and with the idea of a reference system.

First and foremost, Bohr insisted that when a measuring apparatus, or parts of it, is treated quantum mechanically, there must then always be a different system which is treated classically. Formally, this is reflected in the contextuality of the wave function, that is, that such a wave function can be ascribed only to a system which finds itself in a classically described context. In Bohr's own words, one may therefore treat an apparatus quantum mechanically but:[18]

> ...in each case some ultimate measuring instruments, like the scales and clocks which determine the frame of space-time coordination – on which, in the last resort, even the definitions of momentum and energy quantities rest – must always be described *entirely* on classical lines, and consequently kept outside the system subject to quantum mechanical treatment. [My emphasis]

A way to understand Bohr's requirement is that we need a reference frame to make sense of, say, the position of an electron (in order to establish with respect to *what* an electron has a position). And, by definition, a reference frame has a well-defined position and state of motion (momentum). Thus the reference frame is not subject to any Heisenberg uncertainty, and it *is* in this sense (and in this context) classical. This does not exclude that any given reference system could itself be treated quantum mechanically, but we would then need another – classically described – reference system e.g. to ascribe position (or uncertainty in position) to the former.[19]

---

[18] Bohr (1938), p. 104.
[19] See also Dickson (2004) who argues that this requirement of a classical reference frame is crucial in

I think Bohr's view can be summarized in the claim that any system may in principle be seen as and treated quantum mechanically, but not all systems can be seen as and treated in this way simultaneously.[20] This reading of Bohr's viewpoint might be called restricted quantum universalism. It implies that Bohr rejects epistemological quantum fundamentalism since, in a given experimental context, part of the total system – usually the measuring device – must necessarily be described as classical. But Bohr also rejects ontological quantum fundamentalism if this is taken to mean that objects are ultimately represented correctly by a wave function. Both because the wave function is symbolic, and because the answer to what an object is, quantum or classical, depends on the context. This idea of contextualism can be seen as analogous to the standard idea of wave-particle duality which holds that whether, e.g., an electron is a wave or a particle depends on the experimental context.

Let me now return to the question of how Bohr saw the relationship between the quantum and the classical. In this regard, Bohr emphasized that quantum mechanics is a rational generalization of classical mechanics, among other reasons because the formal expression of classical mechanics can be derived from quantum mechanics in the limit where all actions are large compared to Planck's constant. Nevertheless, Bohr would most likely have agreed with the way Landau and Lifshitz expressed the relationship between the two theories:[21]

> Quantum mechanics occupies a very unusual place among physical theories: it contains classical mechanics as a limiting case, yet at the same time it requires this limiting case for its own formulation.

## 6. Outlook

So, do we live in a quantum world? According to Bohr, as I understand him, the answer is no: we live in a world that, at the physical level, can be understood only by referring to both quantum and classical mechanics. In this sense, Bohr's view contrasts with a dominant reductionist way of thinking in physics. According to reductionism, less fundamental theories (like classical mechanics) should be reducible, or derived in terms of, more fundamental theories (like quantum mechanics). Such a program is defended for example by string theorists who hope to find a quantum theory, that both describes the ultimate building blocks (strings) of the universe, and that can also be used to derive all other physical theories. But to the extent that string theory, in a reductionist spirit, is seen as a purely quantum mechanical approach to the world, it also faces the measurement problem. This problem and, more generally, the problem of the still lacking explanation for the transition between quantum indeterminacy and classical well-definiteness, also emerge in modern cosmology. Thus many cosmologists are attracted to the idea of a quantum description of the very early universe which, through a so-called inflationary epoch, gives way to a subsequent classical description of matter in the universe. However, in recent studies, Rugh and myself have tried to show that also in cosmology one must always consider and describe parts of the total system

---

Bohr's reply to Einstein, Podolsky and Rosen regarding the incompleteness of quantum mechanics.
[20] Whether this claim can be sustained for all physical properties, that is, whether a quantum treatment can be given for all degrees of freedom of a physical system, is an open question (Rugh and Zinkernagel 2016).
[21] Landau and Lifshitz (1981), p. 3.

(parts of the constituents of the universe) as classical.[22]

I hope these remarks may help to illustrate the continued relevance of, and indicate the strength in, Bohr's view on quantum mechanics.

**Acknowledgements:** This paper is based on an article in Danish published in the 2013 book *Bohr på ny* [Bohr anew] edited by Lone Bruun, Finn Aaserud and Helge Kragh. The main ideas were first presented at the UK and European Meeting on the Foundations of Physics in Aberdeen 2010. I thank audiences in Aberdeen and Copenhagen, as well as the editors of the present proceedings and an anonymous referee, for helpful comments. I also thank Svend E. Rugh for comments and many discussions on Bohr and quantum theory over the years. Financial support from the Spanish Ministry of Science and Innovation (Project FFI2011-29834-C03-02) is gratefully acknowledged.

## Bibliography

Bohr, Niels (1938). "The causality problem in atomic physics." Reprinted in J. Faye and H. J. Folse (1998), eds., *The Philosophical Writings of Niels Bohr, Vol. IV: Causality and Complementarity*. Woodbridge: Ox Bow, 94-121.

Bohr, Niels (1949). "Discussion with Einstein on epistemological problems in atomic physics." Reprinted in *The Philosophical Writings of Niels Bohr – Vol.II, Essays 1932–1957 on Atomic physics and Human Knowledge. Woodbridge: Ox Bow Press, 1987 (originally, Wiley 1958),* 32-66.

Bohr, Niels (1954). "Unity of Knowledge." Reprinted in *The Philosophical Writings of Niels Bohr – Vol.II,* op.cit., 67-82.

Bohr, Niels (1958). "Quantum Physics and Philosophy – Causality and Complementarity." Reprinted in *The Philosophical Writings of Niels Bohr Vol. III, Essays 1958-1962 on Atomic physics and Human Knowledge*. Woodbridge: Ox Bow Press, 1987 (originally, Wiley 1963), 1-7.

Dickson, Michael (2004). "A view from nowhere: quantum reference frames and uncertainty." *Studies in History and Philosophy of Modern Physics* 35, 195–220.

Favrholdt, David and Hebor, Jens (1999). "What is a realistic understanding of quantum mechanics?" Manuscript presented at the 11th International Congress of Logic, Methodology and Philosophy of Science 1999, Cracow, Poland, 15 pp.

Howard, Don (1994). "What makes a classical concept classical? Toward a Reconstruction of Niels Bohr's Philosophy of Physics." In J. Faye, and H. J. Folse, eds., *Niels Bohr and Contemporary Philosophy*. Dordrecht: Kluwer Academic Publishers, 201-230.

Howard, Don (2004). "Who Invented the 'Copenhagen Interpretation'? A Study in Mythology." *Philosophy of Science* 71, 669-682.

Landau, Lev D. and Lifshitz, Evgeny M. (1981). *Quantum Mechanics: Non-Relativistic Theory, Volume 3, Third Edition (Paperback)*. Oxford: Butterworth-Heinemann.

Landsman, Nicolaas P. (2006). "When champions meet: Rethinking the Bohr–Einstein debate." *Studies in History and Philosophy of Modern Physics* 37, 212–242.

Landsman, Nicolaas P. (2007). "Between classical and quantum." In J. Butterfield and J. Earman, eds., *Handbook of the Philosophy of Science Vol. 2: Philosophy of Physics*. Amsterdam: North-Holland, Elsevier, 417-554.

Petersen, Aage (1963). "The Philosophy of Niels Bohr." *Bulletin of the Atomic Scientists* 19, 8-14.

---

[22] Rugh and Zinkernagel (2011).

Rugh, Svend E. and Zinkernagel, Henrik (2011). "Weyl's principle, cosmic time and quantum fundamentalism." In D. Dieks *et al*., eds., *Explanation, Prediction and Confirmation. The Philosophy of Science in a European Perspective*. Berlin: Springer Verlag, 411-424.

Rugh, Svend E. and Zinkernagel, Henrik (2016). "Schrödinger's cat and the conditions for quantum description(s)." In preparation.

Schlosshauer Maximilian and Camilleri, Kristian (2008), "The quantum-to-classical transition: Bohr's doctrine of classical concepts, emergent classicality, and decoherence", arXiv:0804.1609v1 [quant-ph]

von Neumann, John, (1955). *Mathematical Foundations of Quantum Mechanics*, (trans. by Robert T. Geyer from *Mathematische Grundlagen der Quantenmechanik* 1932), Princeton: Princeton University Press.

Zinkernagel, Henrik (2011). "Some trends in the philosophy of physics." *THEORIA: An International Journal for Theory, History and Foundations of Science*, 71, 215-224.

Zinkernagel, Henrik (2016). "Niels Bohr on the wave function and the classical/quantum divide". *Studies in History and Philosophy of Modern Physics* 53, 9-19.